\documentclass[largeformat]{interact}

\usepackage{epstopdf}
\usepackage{subfigure}

\usepackage[numbers,sort&compress,merge]{natbib}
\bibpunct[, ]{[}{]}{,}{n}{,}{,}

\theoremstyle{plain}

\theoremstyle{definition}

\theoremstyle{remark}

\begin{document}


\title{Activity coefficients of individual ions in LaCl$_{3}$ from the II+IW theory}

\author{
\name{M\'{o}nika Valisk\'{o}\textsuperscript{a} and Dezs\H{o} Boda\textsuperscript{a,b}\thanks{CONTACT D. Boda. Email: boda@almos.vein.hu}}
\affil{\textsuperscript{a}Department of Physical Chemistry, University of Pannonia, P.O. Box 158, H-8201 Veszpr\'em, Hungary}
\affil{\textsuperscript{b}Institute of Advanced Studies K\H{o}szeg (iASK), Chernel st. 14. H-9730 K\H{o}szeg, Hungary}
}

\maketitle

\begin{abstract}
We investigate the individual activity coefficients of ions in LaCl$_{3}$ using our theory that is based on the competition of ion-ion (II) and ion-water (IW) interactions (Vincze {\it et al.}, J. Chem. Phys. 133, 154507, 2010).
The II term is computed from Grand Canonical Monte Carlo simulations on the basis of the implicit solvent model of electrolytes using hard sphere ions with Pauling radii.
The IW term is computed on the basis of Born's treatment of solvation using experimental hydration free energies.
The results show good agreement with experimental data for La$^{3+}$. 
This agreement is remarkable considering the facts that (i) the result is the balance of two terms that are large in absolute value (up to $20 kT$) but opposite in sign , and (ii) that our model does not contain any adjustable parameter.
All the parameters used in the model are taken from experiments: concentration dependent dielectric constant, hydration free energies, Pauling radii.
\end{abstract}


\noindent  \textit{This paper is dedicated to honor of the life time achievement of Professor Johann Fischer in the field of statistical mechanics of condensed phases.}

\clearpage


\section{Introduction}
\label{sec:intro}

A well-known experimental fact in the world of electrolytes is that the activity coefficient (the excess chemical potential) depends on the electrolyte concentration non-monotonically; it decreases near infinite dilution according to the Debye-H\"uckel (DH) limiting law \cite{debye-huckel}, goes through a minimum, and increases at high concentrations close to saturation.
Many theoretical works tried to explain this behavior from various empirical modifications \cite{robinson-stokes,bockris-reddy,fawcett-book,abbas-fpe-07,abbas-jpcb-09,fraenkel_mp_2010,liu_cpl_2015,shilov_jpcb_2015} of the DH theory \cite{debye-huckel} to advanced statistical mechanical approaches based on the model of charged hard spheres immersed in an implicit-water dielectric solvent \cite{triolo-1-jpc-76,triolo-2-jpc-78,triolo-3-jcp-77,simonin-jpc-96,simonin-jpcb-97,lu-fpe-93,fawcett-jpc-96,tikanen-bbgs-96,tikanen-jeac-97,lopezperez_jeac_2000,abbas-fpe-07,abbas-jpcb-09,inchekel}.

In most of these papers before 2010, the dielectric constant was either constant (fixed at its value at infinite dilution, e.g., the dielectric constant of water) or changing with concentration but the change in the interaction with the solvent ignored.
A few notable exceptions are the papers of Abbas et al.\ \cite{abbas-fpe-07} and Inchekel et al.\ \cite{inchekel}.
In 2010, we proposed the II+IW theory \cite{vincze-jcp-133-154507-2010,vincze-jcp-134-157102-2011,valisko-jcp-140-234508-2014,valisko-jpcb-119-1546-2015,valisko-jpcb-119-14332-2015} in which we split the excess chemical potential into two components 
\begin{equation}
\mu_{i}^{\mathrm{EX}} = \mu^{\mathrm{II}}_{i} + \mu^{\mathrm{IW}}_{i},
\end{equation} 
where the II and IW terms describe the ion-ion and ion-water interactions, respectively (subscript $i$ refers to an ionic species).
This equation implies the approximation that these two terms can be computed independently.

In this theory we took into account the experimental fact that the dielectric constant of the electrolyte changes with concentration due mainly to dielectric saturation (see Fig.\ \ref{Fig1}) \cite{bockris-reddy,hasted_jcp_1948,hasted_jcp_1958,giese_jpc_1970,barthel_1970,barthel-data,pottel,helgeson_ajs_1981,wei_jcp_1990,barthel_pac_1991,wei_jcp_1992,nortemann_jpca_1997,buchner_jpca_1999}.
If the properties of the surrounding solvent change, the interaction of the ion with this environment also changes.
This is reflected in the IW term.
This is the term that was ignored by many authors  \cite{triolo-3-jcp-77,simonin-jpc-96,lopezperez_jeac_2000,fawcett-jpc-96,tikanen-bbgs-96,tikanen-jeac-97}.
We proposed, similarly to other authors \cite{abbas-fpe-07,inchekel}, that this term can be estimated with the Born-theory of solvation \cite{born}.

As far as the II term is concerned, several possibilities are available. 
It can be calculated with Grand Canonical Monte Carlo (GCMC) simulations \cite{malasics-jcp-132-244103-2010,malasics-jcp-128-124102-2008} as we did in our previous papers and in this work.  
The mean spherical approximation (MSA) can also be used as we demonstrated \cite{vincze-jcp-133-154507-2010}.
Recently, the concept of the concentration-dependent dielectric constant captured the imagination of Shilov and Lyashchenko \cite{shilov_jpcb_2015} who followed H\"uckel  \cite{huckel_pz_1925} (who advocated this concept as early as 1925) and extended the DH theory for this case.
In our comment \cite{valisko-jpcb-119-14332-2015} to the paper of Shilov and Lyashchenko, we have shown that our IW term is equivalent with their solvation term and shown that the extended DH theory gives better results if we use the Born radius in this solvation term.
In another recent paper, the concentration dependent dielectric constant was built into the model in the Poisson-Fermi theory of Liu and Eisenberg \cite{liu_cpl_2015}.

The main message of the II+IW theory is that the non-monotonic behavior of the activity coefficient is the result of the competition of the II and IW terms.
In earlier works, where the dielectric constant was unchanged, this behavior was reproduced by assigning unrealistically large radii to the hard sphere ions.
The non-monotonic behavior in this case was the result of the balance of volume exclusion (hard sphere) and electrostatic terms.

It was said that the increased radius is a ``solvated ionic radius'' that takes solvation into account by including the hydration shell of tightly connected and oriented water molecules around the ion.
We criticized the concept of the ``solvated radius'' in our previous papers \cite{vincze-jcp-133-154507-2010,vincze-jcp-134-157102-2011,valisko-jcp-140-234508-2014,valisko-jpcb-119-1546-2015,valisko-jpcb-119-14332-2015} and pointed out that important configurations corresponding to cations and anions in contact are excluded from the statistical sample with this artificial concept.
The enlarged solvated radius is primarily useful when we talk about the interactions of ions with water.
That is why the Born-radius is used in the calculation of the IW term (this is equivalent with fitting the IW term to experimental hydration free energies).
In the II term, however, we use the Pauling radii \cite{pauling} of the ``bare'' ions.

Similarly to our previous papers \cite{valisko-jpcb-119-1546-2015,valisko-jpcb-119-14332-2015}, we also present results for activity coefficients of individual ions in this work. 
We are aware of the turbulence about the measurability (and even the existence) of the single ion activities within the electrochemistry community.
There is a recent debate in the literature involving Malatesta \cite{malatesta_jsc_2000,malatesta_fpe_2005,malatesta_fpe_2006,malatesta_fpe_2010,malatest_ces_2010}, Zarubin \cite{zarubin_jctd_2011,zarubin_jctd_445,zarubin_jctd_451}, and the group of Vera and Wilczek-Vera \cite{wilczekvera_aiche_2004,wilczekvera_fpe_2005,wilczekvera_fpe_2006a,wilczekvera_fpe_2006b,wilczekvera_iecr_2009,wilczekvera_ces_2010,wilczekvera_fpe_2011,vera_jctd_442,wilczekvera_jctd_449,wilczekvera_jct_2016}.
We do not discuss this issue here; the reader is directed to our previous paper \cite{valisko-jpcb-119-1546-2015}.
We just use the single-ion activity coefficients published by the Vera--Wilczek-Vera group \cite{wilczekvera_aiche_2004}  and Hurlen \cite{hurlen_acssa_1983} for comparison with our theoretical results.
When judging accuracy of theories, however, the reader should be aware that experiments for the determination of single-ion activities are uncertain and subjects of unknown errors hidden in the junction potential.

We emphasize that the II+IW theory in its present form does not contain any adjustable parameters.
All the results shown in this work have been obtained using experimental hydration energies, experimental concentration dependent dielectric constant, and the Pauling radii.
It is quite remarkable that the theory in its bare form can reproduce the non-monotonic behavior.
It especially works well for 2:1 electrolytes, where the excess chemical potentials (and especially the II and IW terms) vary in a wide range measured in $kT$.
We believe that it is more appropriate to build the competition of opposing forces into the theory through additive free energy terms.
The fact that the sum of a large positive (IW) and a large negative (II) term estimates the EX term so well justifies using the concentration dependent dielectric constant as a link between these two terms. 

In this paper, we make a step further and report results for a 3:1 electrolyte, LaCl$_{3}$, for which we managed to find all the necessary experimental data.
The range in which the balance of the IW and II terms produces the qualitatively correct behavior of the EX term is even wider, $\sim 40 kT$ in the case of La$^{3+}$. 

Studying concentrated electrolytes containing multivalent ions has fundamental importance for many reasons.
Concentrated electrolytes tend to be present in confined systems.
The selectivity filter of ionic channels, for example, can contain crowded ions attracted by the charged groups of the protein.
It is a common view that the electrolyte inside the pore has a reduced dielectric constant compared to the bulk solution \cite{boda-jcp-139-055103-2013}.
Divalent ions play a special role in biology \cite{boda-arcc-2014}. 
They are general messenger particles due to their low concentration and double charge.
Energy storage in electrochemical double layer capacitors has a promising prospect when a superconcentrated electrolyte is present in narrow pores \cite{pean_jacs_2015}.
How ions behave under such circumstances is a relevant question especially that the confined systems are in equilibrium with bulk solutions.
The correct description of bulk solutions is, therefore, necessary to describe the complete system properly.

\section{The II+IW theory}
\label{sec:theory}

The individual activity coefficient, $\gamma_{i}$, in an electrolyte solution describes the deviation from ideality through the excess chemical potential
\begin{equation}
\mu_{i}^{\mathrm{EX}} =  kT \ln \gamma_{i}
\end{equation}
that is defined by
\begin{equation}	
\mu_{i}=\mu_{i}^{0}+kT\ln c_{i}+\mu_{i}^{\mathrm{EX}} ,
\end{equation} 
where $\mu_{i}$ is the chemical potential of species $i$, $c_{i}$ is its concentration, $\mu_{i}^{0}$ is a reference chemical potential independent of the concentration, $k$ is Boltzmann's constant, and $T$ is the temperature (298.15 K in this work).
 The reference point is chosen in such a way that $\lim_{c\rightarrow 0}\mu_{i}^{\mathrm{EX}} = 0$, where $c$ is the salt concentration \cite{lewis1923}.
 The salt concentration is defined as $c=c_{+}/\nu_{+}=c_{-}/\nu_{-}$ with $\nu_{+}$ and $\nu_{-}$ being the stoichiometric coefficients of the cation and the anion in a simple electrolyte with the stoichiometry 
 \begin{equation}
  \mathrm{C}_{\nu_{+}} \mathrm{A}_{\nu_{-}} \rightleftharpoons  \nu_{+} \mathrm{C}^{z_{+}} + \nu_{-} \mathrm{A}^{z_{-}},
 \end{equation} 
 where C and A refer to cations and anions, while $z_{+}$ and $z_{-}$ are the valences of the ions.
The mean activity coefficient is defined as
\begin{equation}
 \gamma_{\pm} = \gamma_{+}^{\nu_{+}/\nu}\gamma_{-}^{\nu_{-}/\nu} ,
\end{equation} 
where $\nu=\nu_{+}+\nu_{-}$.
Accordingly, the mean excess chemical potential is computed as
\begin{equation}
 \mu_{\pm}^{\mathrm{EX}} = \dfrac{\nu_{+}}{\nu} \mu_{+}^{\mathrm{EX}} + \dfrac{\nu_{-}}{\nu} \mu_{-}^{\mathrm{EX}}.
\label{eq:mixmean}
\end{equation} 
The mean quantities, $\gamma_{\pm}$ and $\mu_{\pm}^{\mathrm{EX}}$, can be measured accurately \cite{robinson-stokes,bockris-reddy,fawcett-book}.

\subsection{Calculation of the II term}
\label{sec:calc-II}

\begin{table}[b]
\renewcommand{\arraystretch}{1}
\begin{center}
\begin{tabular}{l|clcl}
Ion & \hspace{0.2cm}$z_{i}$\hspace{0.2cm} & $R_{i} /${\AA} & $\Delta G_{i}^{\mathrm{s}}$/kJmol$^{-1}$ & $R_{i}^{\mathrm{B}} /${\AA}  \nonumber \\
\hline
Na$^{+}$ & 1 & 0.95 & -424 & 1.62   \\
Ca$^{2+}$ & 2 & 0.99 &-1608 & 1.71   \\
La$^{3+}$ & 3 & 1.05 & -3145 & 1.96   \\
Cl$^{-}$ & -1 & 1.81 & -304 & 2.26  \\
\end{tabular}  
\caption{Experimental parameters of ions studied in this work: the valence, $z_{i} $, the Pauling radius \cite{pauling}, $R_{i}$, the hydration Gibbs free energy, $\Delta G_{i}^{\mathrm{s}}$ (taken from Fawcett for Na$^{+}$, Ca$^{2+}$, and Cl$^{-}$ \cite{fawcett-book}, and from Marcus for La$^{3+}$ \cite{Marcus_jcsft_1991}), and the Born radius, $R_{i}^{\mathrm{B}}$ (computed from $\Delta G_{i}^{\mathrm{s}}$ on the basis of Eq.\ \ref{eq:bornw} with $\epsilon_{\mathrm{w}}=78.37$).
}
\label{tab:ions}
\end{center}
\end{table} 

\begin{figure}[t]
 \begin{center}
\resizebox{7cm}{!}{\includegraphics*{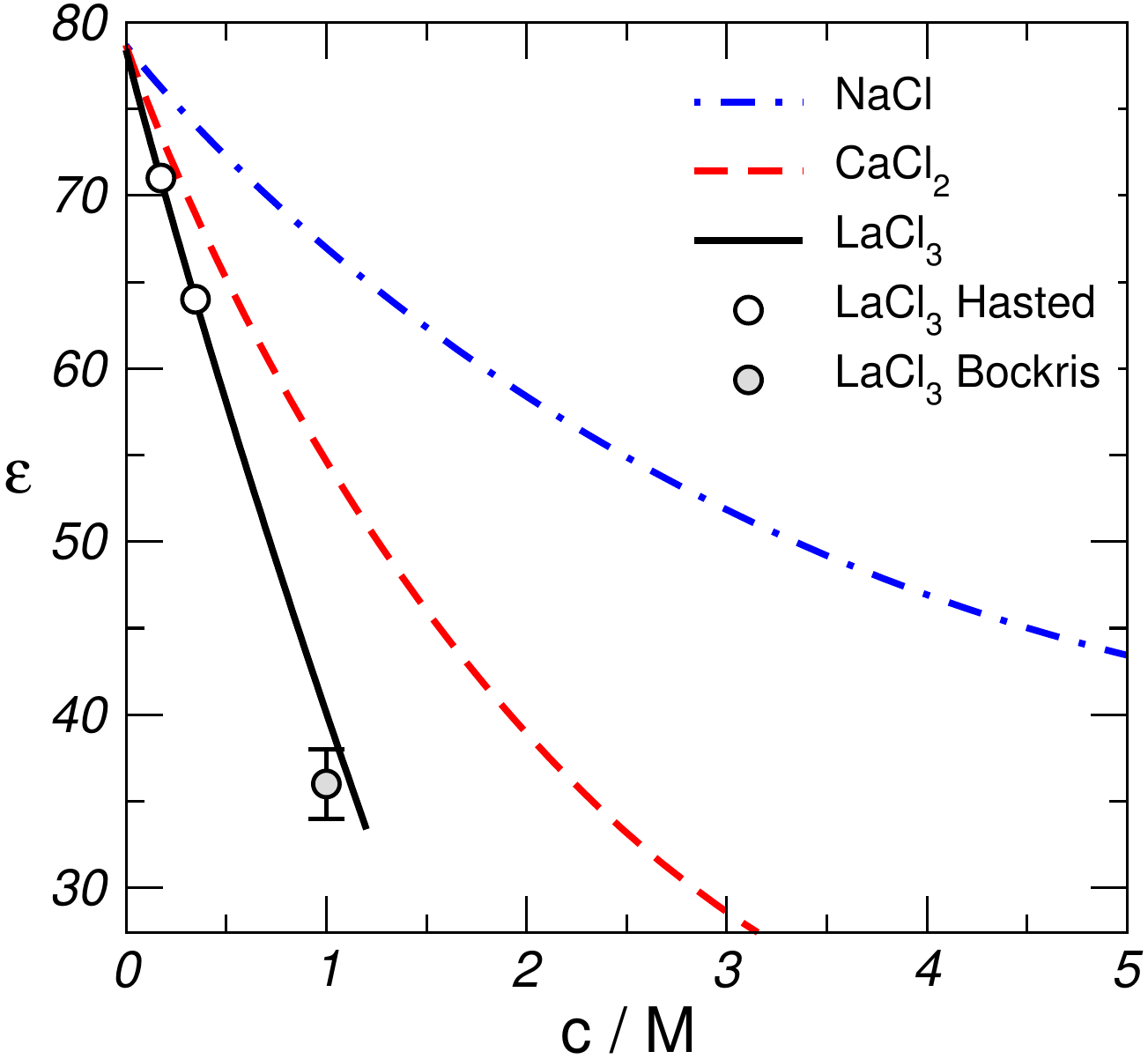}}
 \end{center}
\caption{\small Concentration dependent dielectric constant from measurements. The equations of the fitted curves are $\epsilon (c) = 78.37 - 15.45\, c + 3.76\, c^{3/2} $ for NaCl \cite{buchner_jpca_1999}, $\epsilon (c) = 78.37 - 34\, c + 10\, c^{3/2} $ for CaCl$_{2}$ \cite{barthel_1970,barthel-data}, and $\epsilon (c) = 78.37 - 46.48\, c + 8.21\, c^{3/2} $ for LaCl$_{3}$, where $c$ is the electrolyte (cation) concentration. The equation for LaCl$_{3}$ has been fitted to the data of Hasted et al. \cite{hasted_jcp_1948}, while the point of Bockris  \cite{bockris-reddy} was found afterwards.}
\label{Fig1}
\end{figure}

The II term is calculated on the basis of the Primitive Model (PM) of electrolytes, where the ions are modeled as charged hard spheres, while the solvent is modeled as a dielectric continuum with a dielectric constant $\epsilon(c)$.
The pair-potential describing the II interaction can be given as
\begin{equation}
 u^{\mathrm{PM}}_{ij}(r)
=\left\{
        \begin{array}{ll}
    \infty & \; \mbox{for} \; \;  r<R_{i}+R_{j}\\
        \dfrac{z_{i}z_{j}e^{2}}{4\pi \epsilon_{0}\epsilon(c) r} & \; \mbox{for} \; \; r \geq R_{i}+R_{j}  ,
        \end{array}
        \right. 
\label{eq:pm}
\end{equation} 
where $R_{i}$ is the radius of ionic species $i$, $\epsilon_{0}$ is the permittivity of vacuum, and $r$ is the distance between the ions.
The radius is the Pauling radius in this work, although other choices are also possible \cite{valisko-jcp-140-234508-2014}.
The Pauling radii and other data are collected in Table \ref{tab:ions}.

The experimental dielectric constant, $\epsilon (c)$, is shown in Fig.\ \ref{Fig1} as a function of concentration.
The equations fitted to the experimental data are found in the caption of the figure.
Note that these data are also to be considered with restrictions, because they are obtained from extrapolations to zero frequency using impedance measurement data. 
For example, the results reported by  Barthel {\it et al.} \cite{barthel_1970,barthel-data} from the 1970s for NaCl are quite different from the data of Buchner {\it et al.} \cite{buchner_jpca_1999} from 1999, who used lower frequencies (200 MHz vs.\ 1 GHz). 
In our previous work, however, we showed that small changes in the $\epsilon (c)$ function has little effect on the activities, because these changes have effects of opposite signs in the II and IW terms that tend to cancel each other.
In any case, this effect is smaller than the accuracy of the theory.

We used the the Adaptive-GCMC simulation method of Malasics et al.\ \cite{malasics-jcp-132-244103-2010,malasics-jcp-128-124102-2008} to determine the II excess chemical potentials of individual ions that correspond to prescribed concentrations.
It is an iterative algorithm based on repeated GCMC simulations.
The core of the algorithm uses the assumption that the $kT\ln c_{i}$ term changes more  than the $\mu_{i}^{\mathrm{EX}}$ term over a little change in the state point.
In the case of large ($c>1$ M) concentrations of LaCl$_{3}$, this assumption does not hold, because of the strong correlations between ions (due to  the small dielectric constant and the presence of the trivalent cation that correlates strongly with the anions).
For larger concentrations, therefore, we used the simple method of performing several GCMC simulations on a grid and determining the chemical potentials from interpolation. 

\subsection{Calculation of the IW term}
\label{sec:iw}

The IW term is computed from 
\begin{equation}
 \mu_{i}^{\mathrm{IW}}(c) = \Delta G_{i}^{\mathrm{s}} \dfrac{\epsilon(c)-\epsilon_{\mathrm{w}}}{\epsilon(c) \; (\epsilon_{\mathrm{w}}-1)} ,
\label{eq:iwscaled}
\end{equation} 
where $\epsilon_{\mathrm{w}}=78.37$ is the dielectric constant of water.
This equation contains only experimental data ($\Delta G_{i}^{\mathrm{s}}$ and $\epsilon(c)$) and is obtained from Born's treatment of solvation \cite{born}.
It is equivalent with equation 
\begin{equation}
\mu_{i}^{\mathrm{IW}} (c) = \dfrac{z_{i}^{2}e^{2}}{8\pi\epsilon_{0} R^{\mathrm{B}}_{i}} \left( \dfrac{1}{\epsilon (c)} -\dfrac{1}{\epsilon_{\mathrm{w}}} \right) .
\label{eq:deltaU}
\end{equation} 
Equation \ref{eq:iwscaled} is obtained by writing up the Born-equation for infinite dilution 
\begin{equation}
 \Delta G_{i}^{\mathrm{s}} = \dfrac{z_{i}^{2}e^{2}}{8\pi \epsilon_{0} R^{\mathrm{B}}_{i}} \left( \dfrac{1}{\epsilon_{\mathrm{w}}} -1\right) ,
\label{eq:bornw}
\end{equation}
and eliminating $R^{\mathrm{B}}_{i}$ (the Born-radius, see Table \ref{tab:ions}) from Eqs.\ \ref{eq:deltaU} and \ref{eq:bornw}.
The main point of Eq.\ \ref{eq:iwscaled} is to describe the $\epsilon (c)$-dependence of the IW term in a way that it reproduces the experimental hydration free energies in the $c\rightarrow 0$ limit.

The fact that the ions have different radii in the II and IW calculations should not bother the reader.
The Pauling radii are actual physical radii of the charged hard spheres present in the particle simulations.
The Born radius, on the other hand, is not a physical radius, but rather an effective parameter with which we can make the results of a simple theory match the experimental hydration free energy.

\section{Results}
\label{sec:results}
 
\begin{figure}[t]
 \begin{center}
\resizebox{7cm}{!}{\includegraphics*{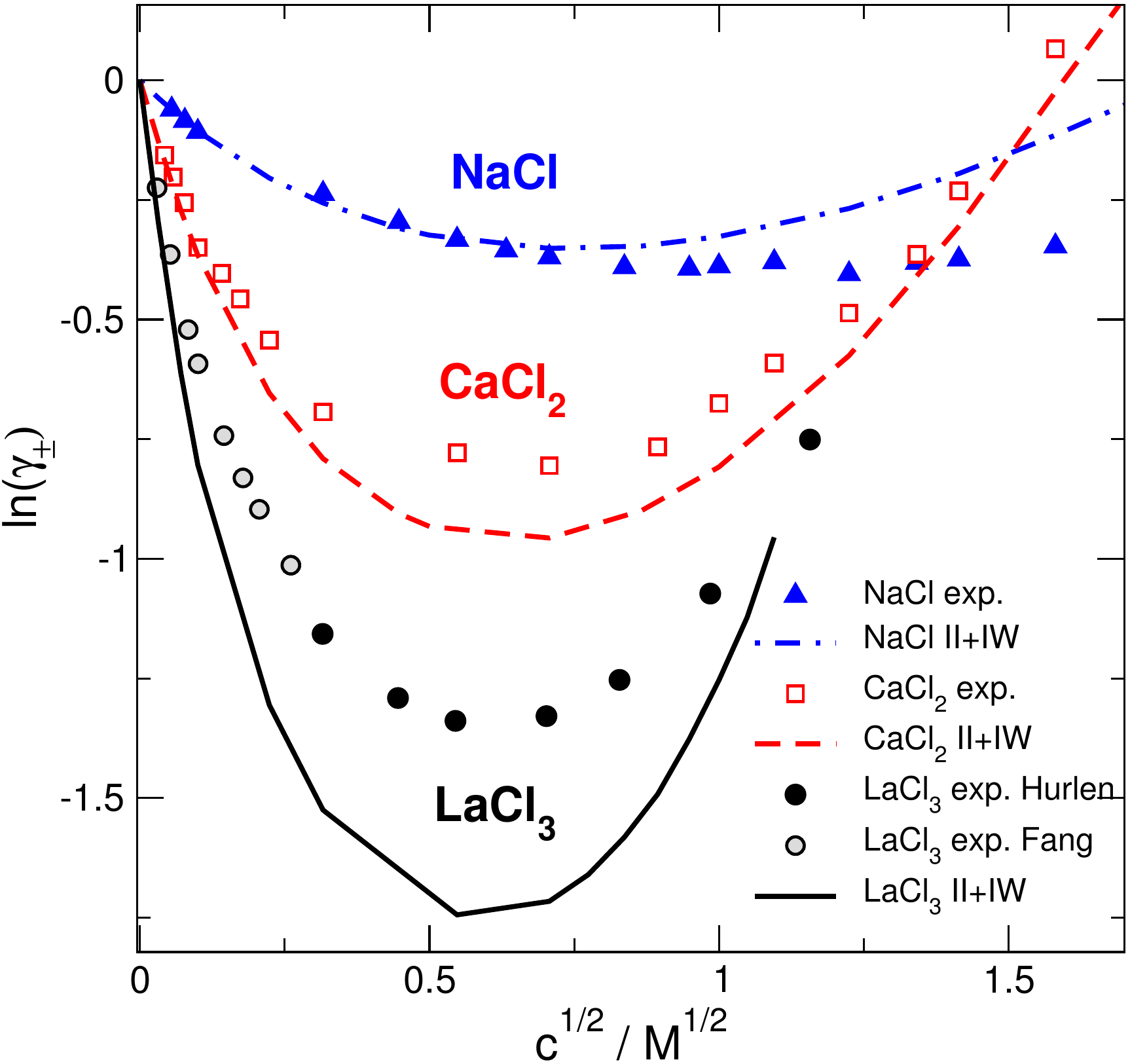}}
 \end{center}
\caption{\small Mean activity coefficients for NaCl, CaCl$_{2}$, and LaCl$_{3}$ as obtained from experiments and the II+IW theory. Experimental data are taken from the following sources: from Wilczek-Vera et al.\ \cite{wilczekvera_aiche_2004} for NaCl, from several works \cite{shedlovsky_jacs_1937,mcleod_jacs_1946,mussini_jced_1971,wilczekvera_aiche_2004} for CaCl$_{2}$, and from Hurlen \cite{hurlen_acssa_1983} and Fang et al. \cite{fang_fpe_2010} for LaCl$_{3}$.}
\label{Fig2}
\end{figure} 

It is reasonable to present the results for the 3:1 electrolyte together with data for 2:1 and 1:1 systems. 
We chose to present results for NaCl and CaCl$_{2}$ taken from our previous paper \cite{valisko-jpcb-119-1546-2015}.
Results for other 1:1 and 2:1 electrolytes can be found in that work.
The data for NaCl and CaCl$_{2}$ represent these systems well as far as the order-of-magnitude comparison between 1:1, 2:1, and 3:1 systems is concerned.

Figure \ref{Fig2} shows the results for the mean activity coefficients.
Note that we plot the logarithm that is just the excess chemical potential in $kT$ unit: $\ln \gamma_{\pm}=\mu_{\pm}^{\mathrm{EX}}/kT$.
The II+IW theory reproduces the non-monotonic behavior qualitatively for all cases without fitting.

\begin{figure}[t]
\begin{center}
\includegraphics*{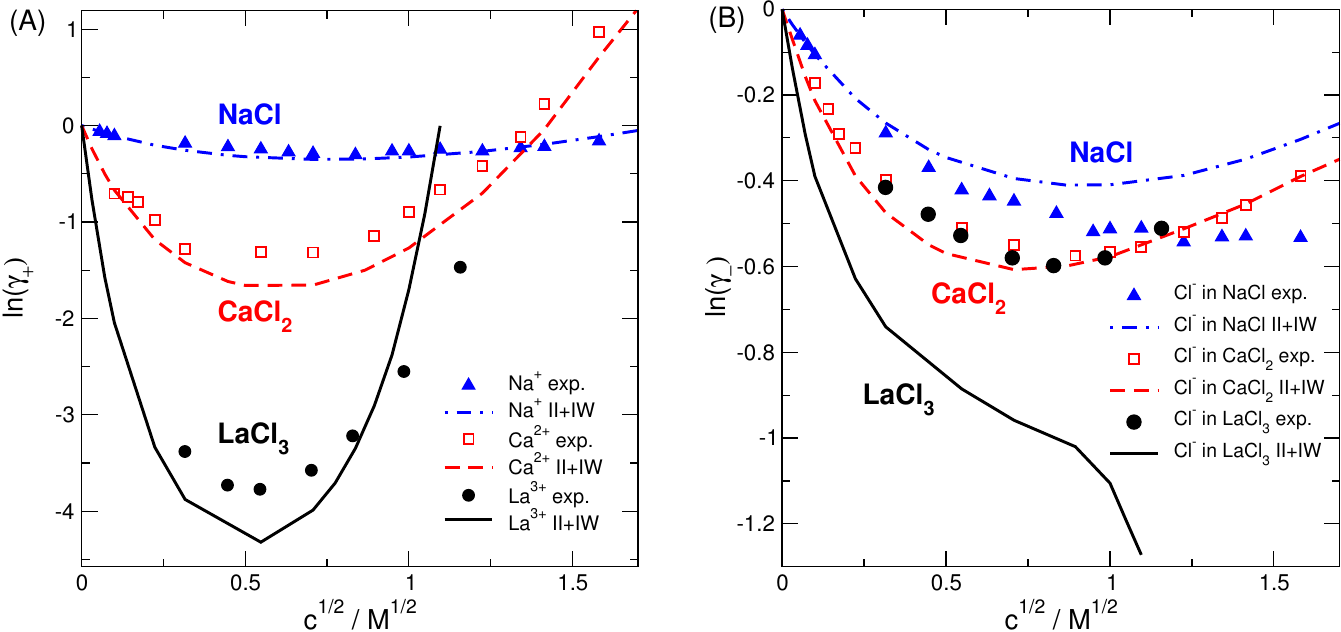}
\end{center}
\caption{\small Individual activity coefficients of (A) the cations and (B) the anions in NaCl, CaCl$_{2}$, and LaCl$_{3}$ as obtained from experiments and the II+IW theory. Experimental data are taken from Wilczek-Vera et al.\ \cite{wilczekvera_aiche_2004} for NaCl and CaCl$_{2}$, and from Hurlen \cite{hurlen_acssa_1983} for LaCl$_{3}$.}
\label{Fig3}
\end{figure}

The minimum of the $\ln \gamma _{\pm}(c)$ curve is deeper for electrolytes containing multivalent ions. 
It is deeper for CaCl$_{2}$ and even deeper for LaCl$_{3}$.
This behavior can be understood by looking at the activity coefficients of the individual ions, especially the cations.
Panels A and B of Fig.\ \ref{Fig3} show the results for cations and anions, respectively.

The excess chemical potentials of Ca$^{2+}$ and La$^{3+}$ vary in a wider range than those of Cl$^{-}$ due to their larger valence.
The cation term is responsible for the more negative values of the mean excess chemical potentials.
The minimum for La$^{3+}$ is about $-4kT$.
Although the excess chemical potential of Cl$^{-}$  appears in the mean with a larger weight (see Eq.\ \ref{eq:mixmean}), this term is restricted to a more narrow range, typically between 0 and $-1kT$.

In what case can we consider the quantitative agreement good? 
Looking at the figures (Figs.\ \ref{Fig2} and \ref{Fig3}), the picture is quite diverse (it gets even more diverse if we look at all the electrolytes published in Ref.\ \cite{valisko-jpcb-119-1546-2015}).
In some cases, the agreement is excellent (Ca$^{2+}$ and Cl$^{-}$ in CaCl$_{2}$, Na$^{+}$ in NaCl). 
In some cases, there is a systematic deviation (Cl$^{-}$ in NaCl and in LaCl$_{3}$).
In some cases, the relative deviation is considerable, but we feel the qualitative agreement appealing (La$^{3+}$).

Interestingly, the cases, when the absolute value of $\ln \gamma_{i}$ is small (below $1$) seem to be the difficult cases.
These are the cases of the monovalent ions (Na$^{+}$ and Cl$^{-}$).
The two cases that are the most interesting for us are those of the multivalent cations (Ca$^{2+}$ and La$^{3+}$).
In these cases the energies involved are so large that we value the qualitative agreement and we are not bothered by the quantitative disagreement too much.

\begin{figure}[t]
 \begin{center}
\includegraphics*{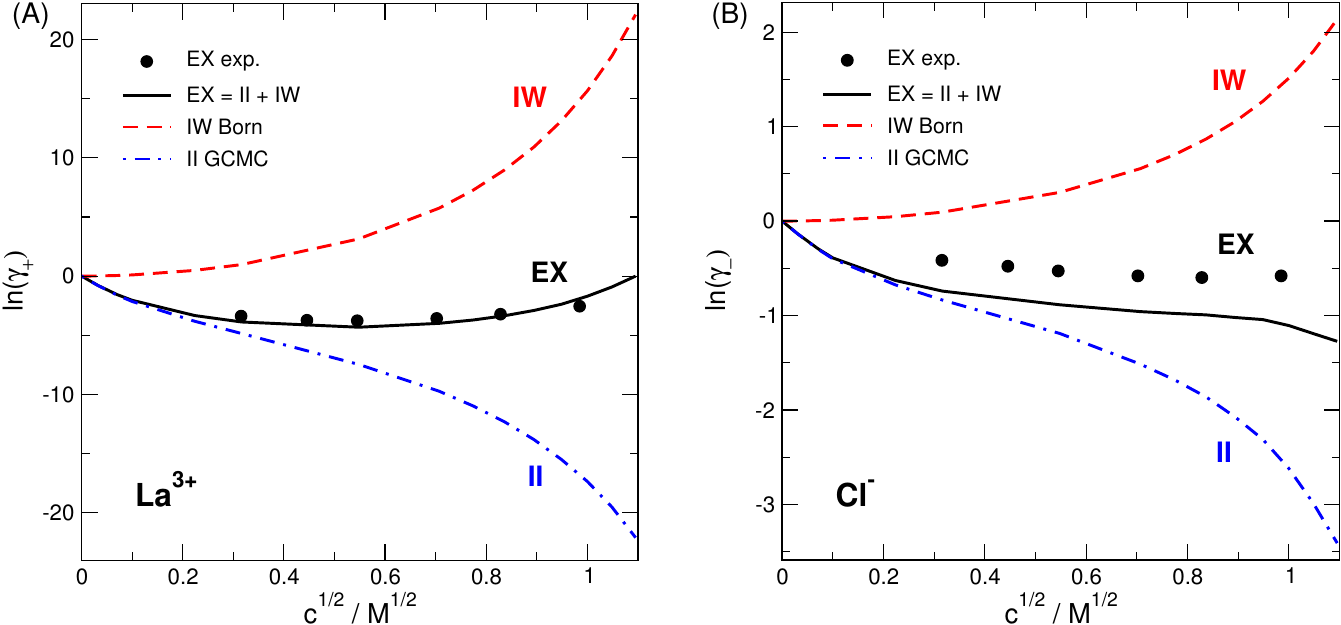}
 \end{center}
\caption{\small The II and IW components of the individual activity coefficients (denoted by EX) of (A) La$^{3+}$  and (B) Cl$^{-}$ in LaCl$_{3}$ as obtained from the II+IW theory. The II component has been obtained from GCMC simulations using the Pauling radii (Table \ref{tab:ions}) for the charged hard spheres, while the IW component has been obtained from the Born-theory using experimental hydration free energies (Table \ref{tab:ions}) in Eq.\ \ref{eq:iwscaled}. Experimental data are taken from Hurlen \cite{hurlen_acssa_1983}.}
\label{Fig4}
\end{figure}

The quantitative explanation of all these ``feelings'' and ``considerings'' is that the non-monotonic behavior is the result of the balance of two competing effects.
In our model, these competing effects are the II and IW interactions.
If we increase concentration, $\epsilon(c)$ decreases (Fig.\ \ref{Fig1}) and the Coulomb interactions get less screened (Eq.\ \ref{eq:pm}).
At larger concentrations, the ions are closer to each other on average.
Therefore, the II term becomes more and more negative as $c$ increases.
The IW term, on the other hand, increases with increasing $c$, because the dielectric environment gets less and less favorable for the ion as $\epsilon (c)$ decreases.
At small to moderate concentrations, the II term decreases faster, while at larger concentrations above the minimum point, the IW term catches up and increases faster; hence the non-monotonic behavior.
 
These two balancing terms are plotted in panels A and B of Fig.\ \ref{Fig4} for La$^{3+}$ and Cl$^{-}$, respectively, in LaCl$_{3}$.
This figure explains why the qualitative agreement for La$^{3+}$ is so valuable (the corresponding figure for Ca$^{2+}$ can be found in our previous papers \cite{valisko-jpcb-119-1546-2015,valisko-jpcb-119-14332-2015}).
The II term goes down to $-20kT$ and the IW term goes up to $20kT$ at concentrations close to saturation.
Their sum (EX) is so close to the experimental value (within $0.5kT$) that we consider the agreement remarkable especially in the light of the fact that our model does not contain any adjustable parameters.
Looking at the results for many 1:1 and 2:1 electrolytes \cite{valisko-jpcb-119-1546-2015}, deviations are usually in the range of a few tenth of $kT$s.
That seems to be a natural limit to the accuracy of our very simplified model.

The situation with Cl$^{-}$ is the opposite. 
Here, the II and IW terms are within $3kT$ and the sum is smaller than $1kT$, so there is a more delicate balance in this case. 
Any error in the II and IW terms separately has a larger impact here, but its effect on the mean is less, because the cation dominates the mean.

\begin{figure}[t]
 \begin{center}
\includegraphics*{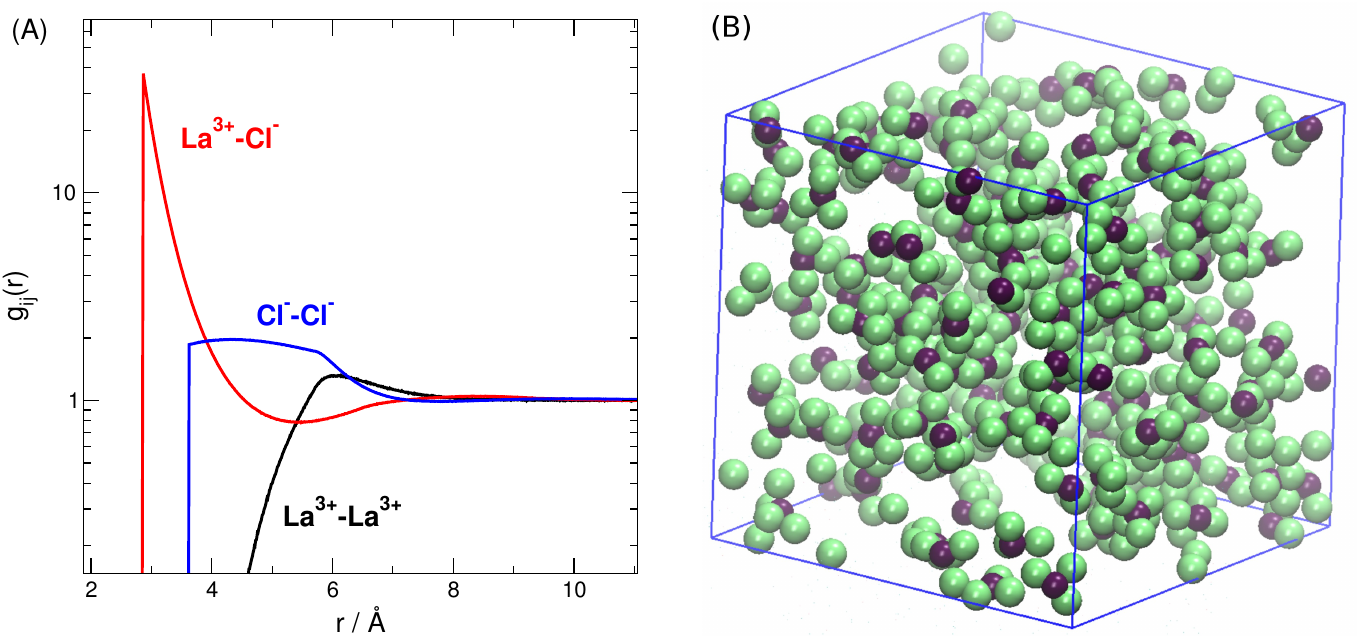}
 \end{center}
\caption{\small (A) Radial distribution functions for $c=1.2$ M obtained from canonical MC simulations. Profiles for lower concentrations are similar but with lower peaks. (B) A snapshot from the simulation.}
\label{Fig5}
\end{figure}
 
For LaCl$_{3}$, results are shown only for concentrations up to 1.2 M. 
At larger concentrations, close to 1.2 M, the interactions between ions become stronger due partly to the reduced dielectric constant, partly to the shorter average distance between the ions.
Because of the strong interactions, successful insertions and deletions of ions (especially, La$^{3+}$) are very rare, and sampling in GCMC simulations becomes inadequate above 1.2 M.
Recent studies using molecular dynamics simulations, spectral graph theory, and various experimental techniques imply the presence of ion aggregates from ion pairs to networks in NaCl and KSCN \cite{choi1-jcp-2014,choi2-jcp-2014,choi3-jcp-2015,choi4-jcp-2015,choi5-jcp-2016,choi6-jcp-2016}.
There is no reason to be surprised that strongly correlated ions in ordered structures appear as we approach the solubility limit \cite{alejandre_pre_2007,alejandre_jcp_2009}.
The ions of the electrolyte, however, can be considered solvated even if these structures are present.
The radial distribution functions of Fig.\ \ref{Fig5}A for $c=1.2$ M show a fluid-like behavior at long distance, while large first peaks indicate the increased probability of ions forming aggregates.
Their nature is revealed by the snapshot shown in Fig.\ \ref{Fig5}B.
These aggregates, however, are not solid structures, but momentary groups of ions that are forming and breaking up during thermal motion.
The formation of these is closely related to the disruption of the hydrogen-bond network of water, so the simulation study of the phenomena is probably beyond the scope of the implicit solvent framework used here. 
The solubility of electrolytes, a closely related problem, has been thoroughly investigated by several research groups \cite{lisal_jpcb_2005,sanz_jcp_2007,moucka_jpcb_2011,aragones_jcp_2012,moucka_jpcb_2012,benavides_jcp_2016} using explicit-water models.

Summarized, we feel that the relatively good agreement of the individual excess chemical potential with experiments for multivalent cations (results for Mg$^{2+}$ and Ba$^{2+}$ are found in our previous work \cite{valisko-jpcb-119-1546-2015}) is a strong evidence that supports our theory.
In these cases, the absolute values of the competing II and IW terms are so large that the error in the sum is dwarfed by them.

\section*{Acknowledgements}
 
We gratefully acknowledge the financial support of the Hungarian National Research Fund (OTKA NN113527) in the framework of ERA Chemistry.


\end{document}